\newcommand {\fd} {\rightarrow}

\newcommand {\la} {\langle}
\newcommand {\ra} {\rangle}
\newcommand {\bay} {\begin{array}}
\newcommand {\eay} {\end {array}}
\newcommand {\be}{\begin{equation}\noindent}
\newcommand {\ee}{\end{equation}\noindent}
\newcommand {\ba}{\begin{eqnarray}}
\newcommand {\ea}{\end{eqnarray}}
\newcommand {\bd}{\begin{displaymath}}
\newcommand {\ed}{\end{displaymath}}
\newcommand {\bi}{\begin{itemize}}
\newcommand {\ei}{\end{itemize}}

\newcommand {\formu}[1] {(\ref{#1})}

\documentclass[12pt]{article}

\usepackage{graphicx}
\input english.sty
\begin{document}
\title{Testing a Variational Approach to Random Directed Polymers}
\author{Andrea Pagnani\footnote{\tt{andrea.pagnani@roma2.infn.it}}\\
{\small Dipartimento di Fisica, Universit\`a di Roma}
{\small {\em Tor Vergata} }\\
{\small \ \  Via della Ricerca Scientifica 1, 00133 Roma (Italy)}}
\maketitle
\begin{abstract}
The one dimensional direct polymer in random media model is
investigated using a variational approach in the replica space.
We demonstrate numerically that the stable point is a maximum and the
corresponding statistical properties are for the delta correlated
potential in good agreements with the known analytic solution. In the
case of power-law correlated potential two regimes are recovered: a
Flory scaling dependent on the exponent of the correlations, and a
short range regime in analogy with the delta-correlated potential
case. 
\end{abstract}
\section{Introduction}
A challenging open problem in statistical mechanics is the
behavior of fluctuating random manifolds with quenched disorder (for a
review see \cite{HALPIN_ZHANG}).
New insight has been obtained in the framework of replica formalism
\cite{KARDAR_87} and, at least at the level of the tree approximation
(Hartree-Fock) which is the exact solution in the limit of
infinitely many dimensions, the problem has been hopefully solved
\cite{RANDOM_MAN}. Nevertheless the approximation turned out to be
maybe too drastic for take in account the behavior in all dimension
and for different structures of correlation of the quenched noise. 
For instance the prediction for the wandering exponent $\zeta$ of 
one-dimensional direct polymers in
random media (DPRM) are wrong even in the case of short range
correlations of the random potential.\par
In the following we will restrict to DPRM, which due to their
one-dimensional structure are simpler then random manifolds with
generic dimension.\par
After this short introduction we will present some details on the
model and we will define some basic mathematical tools in order to
introduce in section $3$ the variational approach developed in
\cite{PARISI_TESI} and reformulating it in a suitable form to
implement a numerical simulation.
Section $4$ is divided in two parts: in the first one  
we check the agreement of our results with
Kardar's analytic solution \cite{KARDAR_87} for the delta-correlated
potential and in the second part we study the power-law correlated
potential for different values of its exponent. Two different scaling
regimes are recovered: a short-range one analogous to the
delta-correlated potential, and a Flory like regime in wich the
wandering exponent depends from the power-law correlation exponent in
agreement with the one loop renomalization group analysis presented in
\cite{MHKZ}. Finally in section $5$  we present comments and
perspectives.
\section{The Model}
Knowledge of end point probability distribution in DPRM is of
fundamental importance. In principle it is always possible to determine
this probability distribution $P_{t}[{\bf x}]$ starting from the
explicit structure of the Hamiltonian of the system which depends on
the space coordinate set ${\bf x}$, the time $t$, and the random
potential $\eta$:
\be
H[{\bf x}]=\int dl\; \left\{\frac{1}{2}\frac{d{\bf x}}{dl}+\eta({\bf
x},t)\right\}
\label{Hamiltonian}
\ee
The related partition function satisfies the imaginary time
Schr\"{o}dinger equation: 
\be
\frac{\partial Z({\bf x},t)}{\partial t}=
\frac{1}{2}\nabla^{2} Z({\bf x},t)+\eta({\bf x},t)Z({\bf x},t)	
\label{Scrotinger}
\ee
The polymer's end point probability distribution function $\rho({\bf
x},t)$ is readily obtained by:
\be
\rho({\bf x},t)=\frac{Z({\bf x},t)}{\int d^{d}{\bf y}\;Z({\bf y},t)}
\label{RHO}
\ee
We want to implement a variational approach in order to determine a
functional of the probability $P_{t}\left[\rho({\bf x},t)\right]$
which allows direct calculation of the quenched noise average of
a generic observable $O$ by the functional integration:
\be
\overline{O[\rho({\bf x},t)]}=
\int d[\rho]P_t[\rho({\bf x},t)]O[\rho({\bf x},t)]
\label{MEDIE_QUENCHED} 
\ee
Quenched averaging is believed to be a rather difficult problem in
spin-glass and DPRM models. Replica technique  
consists on calculating quenched average of the replicated
partition function. We can therefore introduce 
$Z(\{{\bf x}_a\},t)=
\overline{ Z({\bf x_{1}},t)Z({\bf x_{2}},t)...Z({\bf x_{n}},t) }$
as the solution of the imaginary time Schr\"{o}dinger equation:
\be
\frac{\partial Z(\{{\bf x}_a\},t)}{\partial t}=
\left[\frac{1}{2}\sum_{i=1}^{d}\sum_{a=1}^{n}
\frac{\partial^{2}}{\partial x_{i,a}^{2}}+
\sum_{a<b}^{n} V({\bf x}_{a}-{\bf x}_{b})\right]Z(\{{\bf x}_a\},t)
\label{REPL_SCROT} 
\ee
where $V({\bf x}-{\bf y})=\overline{\eta({\bf x},t)\eta({\bf y},t')}$.
A couple of explicit solutions to \formu{REPL_SCROT} are already
available for $V(x-y)=\delta(x-y)$ \cite{KARDAR_87} and 
$V(x-y)=(x-y)^{2}$ \cite{PARISI_90}, both of them in the one
dimensional case $d=1$. Unfortunately those are the only known
analytical solutions of the problem which in general can be
only studied by mean of renormalization group approximate technique
\cite{MHKZ}.
A particular solution of \formu{REPL_SCROT} can be decomposed over
the eigenstate's set of the Hamiltonian:
\be
Z(\{{\bf x}_{a}\},t)=
\sum_{i}c_{i}\Psi_{i}(\{{\bf x^{0}}_{a}\})
\Psi_{i}(\{{\bf x}_{a}\})e^{-E_{i}t}
\label{EIGEN_DECOMP}
\ee
From the previous expression it is evident that all the excited states
decrease sub-exponentially compared with the ground state, and then
the thermodynamic behavior of the model will be ruled in the
asymptotic regime by the structure of the ground state. In this
regime is possible to calculate for instance:
\be
\overline{\rho({\bf x_{1}})\rho({\bf x_{2}})}=
\int dP[\rho]\;\rho({\bf x_{1}})\rho({\bf x_{2}})\;\propto\;
\int d{\bf x_{3}}d{\bf x_{4}}...d{\bf x_{n}}\;
\psi({\bf x_{1}},{\bf x_{2}},...,{\bf x_{n}})
\label{2_MOMENTO} 
\ee
where $\psi({\bf x_{1}},{\bf x_{2}},...,{\bf x_{n}})$ is the ground
state of \formu{REPL_SCROT} and
$dP[\rho]=\lim_{t\fd\infty}dP_t[\rho]$. \par
Since we know the exact solution of the one dimensional
delta-correlated potential model, a careful analysis of its ground
state structure should be illuminating for further development of
the variational principle. In fact a straight application of the Bethe
ansatz - which turns out to be the exact solution of the 
delta-correlated model - shows that the ground eigenstate is:
\be
\psi(x_{1},x_{2},...,x_{n})=\prod_{a<b}e^{\displaystyle -A|x_{a}-x_{b}|}
\label{BETHE_GROUND}
\ee
with energy:
\be
E^{(n)}=-\frac{n(n^2-1)A^2}{3}
\label{BETHE_ENERGY}
\ee
where $A$ is a constant depending on the parameters of the
Hamiltonian. So, following M\'ezard et al. in
\cite{PARISI_TESI}, in the general case a reasonable choice should be
restrict the space of n-particles test functions to the following:
\be
\psi(x_{1},x_{2},...,x_{n})=\prod_{a<b}f(x_{a}-x_{b})
\label{TEST_GROUND}
\ee
A direct calculation shows that this is tantamount to choosing 
the following Gaussian probability distribution $P_{g}[\phi]$ for 
$\phi({\bf x})\equiv\ln(\rho({\bf x}))$:
\be 
P_{g}[\phi]\propto\exp \left[ -\frac{1}{2}\int d^{d}x d^{d}y\;
\phi({\bf x}) g({\bf x}-{\bf y}) \phi({\bf y}) \right] 
\label{GAUSSIAN_PROB}
\ee
In fact performing a Gaussian integration we obtain:
\be
\int d[\phi]\exp\left[ \sum_{a=1}^{n} \phi({\bf x}_{a})-
\frac{1}{2}\int d^{d}x d^{d}y\;
\phi({\bf x})g({\bf x}-{\bf y})\phi({\bf y}) \right]=
\prod_{a,b} f({\bf x}_{a}-{\bf y}_{b})
\ee
and the following relation holds:
\be
\int d^{d}z\;g({\bf x}-{\bf z})\ln[f({\bf z}-{\bf y})]= 
\delta^{d}({\bf x}-{\bf y})
\label{G_F}
\ee
In the previous one dimensional case $g$ must be chosen in such 
a way that:
\be
\int d x d y\;
\phi( x) g(x -y) \phi( y) = \int d x \frac{1}{2}|\nabla\phi|^2
\ee 
It's now reasonable to guess for our
ground state a functional structure of the following type:
\be
\label{FUNCT_STRUCT}
\psi({\bf x}_{1},{\bf x}_{2},...,{\bf x}_{n})\propto
\int dR[\mu]\;\mu({\bf x}_{1})\mu({\bf x}_{2})...\mu({\bf x}_{n})
\ee
The distribution functions $P[\rho]$ and $R[\mu]$ are related by:
\be
\overline{\rho({\bf x}_{1})\rho({\bf x}_{2})}=
\int dP[\rho]\;\rho({\bf x}_{1})\rho({\bf x}_{2})=
\int dR[\mu]\;\frac{\mu({\bf x}_{1})\mu({\bf x}_{2})}{I[\mu]^{2}}
\label{R_P}
\ee
where $I[\mu]=\int d^{d}x\;\mu({\bf x})$.
\section{The Variational Principle}
Since now we have recast our problem in a (hopefully) suitable form
to implement the variational principle which consists on minimizing
the quadratic form $\la\psi|H|\psi\ra$, where $H$ is the replicated
Hamiltonian operator which appears in the square brackets of
\formu{REPL_SCROT}. In order to obtain normalized eigenstates we have
to enforce this condition by the introduction of a Lagrangian
multiplier. Following this procedure described in \cite{PARISI_TESI}
we easily obtain:
\be
\la\psi|H|\psi\ra=\int dR[\mu]dR[\sigma]\mathcal{H}_{n}[\mu,\sigma]
\label{QUADR_FORM}
\ee
where 
\ba
\mathcal{H}_{n}[\mu,\sigma]&=&\frac{n}{2}\int d^{d}x\;
\sum_{\nu=1}^{d}
\frac{\partial \mu}{\partial x_{\nu}}
\frac{\partial \sigma}{\partial x_{\nu}}I[\mu\sigma]^{n-1}+\\
\nonumber
&&n(n-1)\int d^{d}x d^{d}y \;
\mu({\bf x})\mu({\bf y})V({\bf x}-{\bf y})\sigma({\bf x})\sigma({\bf y})
I[\mu\sigma]^{n-2}
\label{REPL_FREE}
\ea
The normalization in terms of our trial functions is:
\be
\la\psi|\psi\ra=\int dR[\mu]dR[\sigma]\;I[\mu \sigma]^{n}
\label{NORMALIZ}
\ee
In this new representation the Schr\"{o}dinger equation reads:
\be
\int dR[\sigma]\;\mathcal{H}_{n}[\mu,\sigma]=
E_{n}\int dR[\sigma]\;I[\mu \sigma]^{n} 
\label{SCROT_NEW}
\ee
It's now possible to consider the $n\fd 0$ limit of
\formu{SCROT_NEW}. In this limit we can eventually implement the
variational principle in terms of the stationary point of the free
energy functional $F[R]=\int
dR[\mu]dR[\sigma]\, \mathcal{H}[\mu,\sigma]$, and $\mathcal{H}$ is now
defined by:
\ba     
\label{ACCAVARn0}       
        \mathcal{H}[\mu,\sigma]&=&\frac{1}{2}\int d^{d}x 
        \sum_{\nu=1}^{d}\frac{\partial \mu}{\partial x_{\nu}} 
        \frac{\partial \sigma}{\partial x_{\nu}}I[\mu\sigma]^{-1}\nonumber\\
        &+&\int d^{d}xd^{d}y\;\mu({\bf x})\mu({\bf y})
        \sigma({\bf x})\sigma({\bf y})V({\bf x}-{\bf y})I[\mu\sigma]^{-2}
\ea
As far as replica symmetric approximation is concerned, the problem is
recast just in finding the proper $g$-function - as defined in
\formu{GAUSSIAN_PROB} - which is stationary against variations of the
following functional: 
\be
\label{MIO}
        \int dP_{g}[\phi_{1}]dP_{g}[\phi_{2}]
        \mathcal{H}[\exp(\phi_{1}),\exp(\phi_{2})]
\ee
Performing analytic calculation on this functional is hopeless, 
so we are forced to try a numerical approach in order to calculate the
critical exponents of the model.

\section {Numerical Analysis}
\subsection{The delta-correlated potential}
As already mentioned, $1$-dimensional DPRM with delta-correlated
random potential is explicitly solvable and so it is a good testing
ground for our variational approach. For technical reasons it is simpler 
to consider the case in which a lattice spacing is introduced.
The Hamiltonian, taking lattice spacing, $a=1$ is:
\ba
\mathcal{H}[\mu,\sigma]&=&\frac{1}{2}\sum_{i=1}^{N}  
        (\mu_{i+1}-\mu_{i})(\sigma_{i+1}-\sigma_{i}) 
        I[\mu\sigma]^{-1}\nonumber\\
        &+&G \sum_{i,j}^{N} \mu_{i}\mu_{j}
        \sigma_i\sigma_jV_{ij} I[\mu\sigma]^{-2}
\ea
where in the delta-correlated case $V_{ij}=\delta_{ij}$.
The 
continuum limit can be reached sending the lattice spacing to zero
or sending the coupling constant measured in lattice units to
zero. Indeed the dimensionless coupling constant is $G a^3$.
Although the wandering exponent is supposed to be independent from the
lattice spacing, the continuum limit may be interesting as a check to
our computation.\par
In order to implement a numerical integration of \formu{MIO} we need
as a first step to generate the fields $\phi$ which follow the
distribution $P_g[\phi]$. The Fourier representation of this
probability distribution is:
\be
\label{PROB_FOUR}
        P_{g}[\phi]d[\phi]\propto\prod_{k<\Lambda}
        d[\phi_{k}]\exp(g_{k}|\phi_{k}|^{2})
\ee
where $k\in\{(\pi n)/L\}_{n=1,2,...N}\;$, $N$ are equally spaced
intervals in which we consider discretized the field
$\phi(x)\;\;x\in\{0,L\}$ and $\,\phi_k\,$ are  Fourier coefficients of
$\phi$. The integration in Fourier space ``decouples'',  in the sense
that now the probability distribution is the product of the individual
$k$-wave vector probability distributions. \par
Assuming that the wave function on the lattice is the same as in the
continuum (the difference among the two going to zero in the continuum
limit), a little algebra reveals that:
\be
\prod_{a<b} e^{\displaystyle -A|x_{a}-x_{b}|}=
\int d[\phi]\prod_{a=1}^{n}\exp\left\{\phi(x_{a})-\int dx\;
\frac{|\nabla \phi(x)|^2}{4A}\right\}
\label{EXP_TRANS}
\ee
which means that for delta-correlated noise we straightforwardly
obtain:   
\be
dP[\phi] \propto 
\exp\left\{\int dx\;\frac{|\nabla \phi(x)|^2}{2A}\right\}d[\phi] 
\propto 
\prod_{k<\Lambda}\exp\left\{\frac{k^2|\phi_{k}|^2}{2A}\right\}d[\phi_{k}]
\label{PROB_DELTA}
\ee
From the structure of \formu{PROB_DELTA} we can reasonable guess the
following structure for the probability distribution:
\be 
dP[\phi] \propto 
\prod_{k<\Lambda}
\exp\left\{-\frac{\lambda k^{2\alpha}|\phi_{k}|^2}{2}\right\}d[\phi_{k}]
\label{PROB_VAR}
\ee
in which both $\alpha$ and $\lambda$ are to be considered as 
variational parameters related with the fluctuations of $\phi$, since:
\be 
\langle(\phi(x)-\phi(y))^{2}\rangle=
        \int dP_{g}[\phi](\phi(x)-\phi(y))^{2}=\lambda|x-y|^{\alpha}
\label{PHI_FLUCT}
\ee
Comparing \formu{PROB_DELTA} with \formu{PROB_VAR} it's easy to deduce
that the delta-correlated case is obtained setting $\alpha=1$, so we
used this result to a first tuning of the variational parameters. We
have generated statistically uncorrelated random numbers distributed
as \formu{PROB_VAR} with $\alpha=1$.
Back Fourier transforming via FFT algorithms \cite{NUM_RECIP} the
coefficients $\phi_{k}$, we are able to obtain statistically uncorrelated
realizations of $\phi(x)$.
Let $\Gamma$ be the set of $M$ realizations of $\phi(x)$, we have
found a suitable interval of values for $\lambda$ in which:
\be
\frac{1}{M}\sum_{a\in \Gamma}
\left(\phi((x_{a})-\phi(y_{a})\right)^{2}=
C|x-y|\;\;\;\;\; C=O(1)
\label{FLUCT_PHI}
\ee
\begin{figure}[!htb]
  \begin{center}
    \includegraphics[width=0.50\textwidth,height =0.25\textheight]{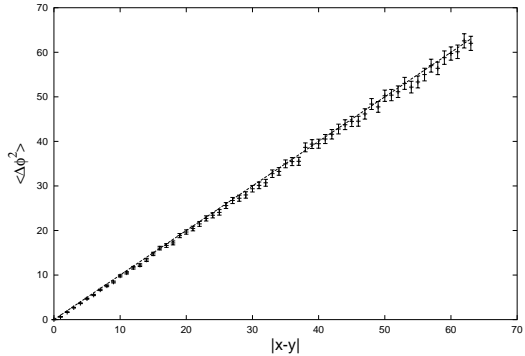}
    \protect\caption{Fluctuations of the field $\phi$ vs. distance 
     averaged over $10^5$ realizations of disorder for
     $(\alpha=1.0,\lambda=1.0)$.  }
    \label{PHI_FLUCT:FIG}
  \end{center}
\end{figure}
The $\phi(x)$ fields built following this prescription are the
configurations used for calculating the value of the free energy
functional $\mathcal{H}$ defined in \formu{ACCAVARn0}. It is obvious
that such a value should depend on the realization of the fields and
so we have to average over a significative number of this
realizations, or more formally, $|\phi_{a}\ra$ and $|\psi_{a}\ra$
being the $a$-th of $M$ realizations, we can compute:
\be 
\langle F\rangle=\frac{1}{M^{2}}\sum_{a,b=1}^{M}\langle\psi_{a}|
\mathcal{H}|\phi_{b}\rangle
\label{BRA_KET}
\ee 
where in this case we use the delta-correlated potential
$V(x-y)=\delta(x-y)$.
It's easier to use intensive variational 
parameters, so we redefine \formu{PROB_VAR} as:
\be
dP_{g}[\phi]=\prod_{k<\Lambda}d[\phi_{k}]\exp\left(-
N^{3/2}\frac{\lambda k^{2\alpha}}{2} \right)
\label{PROG_TRIAL}
\ee
where $\alpha$ and $\lambda$ are new variational parameters.\par
A crucial question now is whether the free energy functional should be
maximized or minimized, since from a strictly formal point of view we are
considering the $n\fd 0$ limit, a regime in which thermodynamic
behavior is not evident at all.
In the general theory proposed in \cite{RANDOM_MAN} it is proved that
Hartree Fock approximation consists in a saddle point on the number of
components of the field going to infinity, and the solution is a
maximum of the functional. 
A possible argument which explains this rather unusual feature of
replica theory is that the Hamiltonian \formu{REPL_SCROT} is
translational invariant, so if we choose the center of mass reference
frame we are left with $n-1$ degree of freedom (a number which is
negative as soon as $n<1$ !). This is the regime in which the stationary
point becomes a maximum of the functional.\par
\begin{figure}[!htb]
\begin{center}
  \begin{tabular}{rl}      
    \includegraphics[width=0.45\textwidth,height =0.25\textheight]{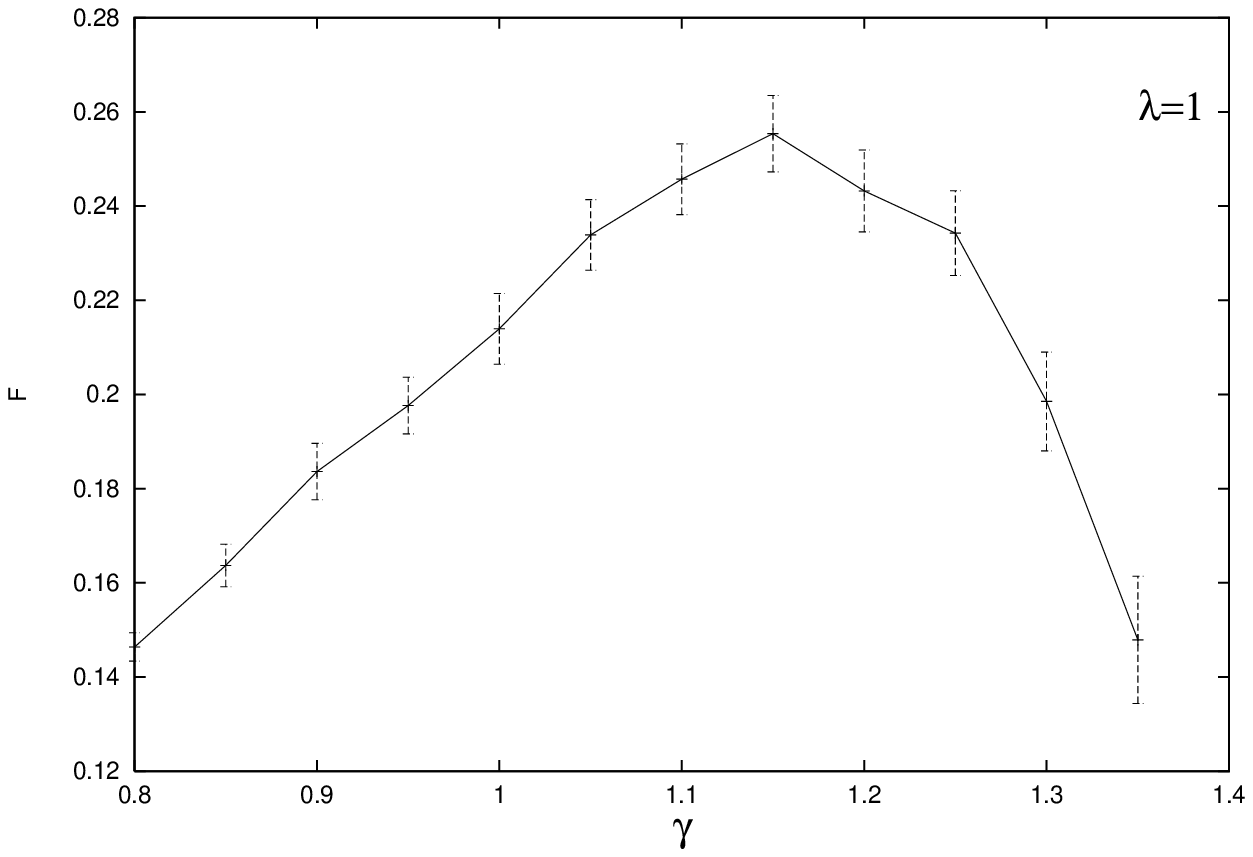}
   	&
    \includegraphics[width=0.45\textwidth,height =0.25\textheight]{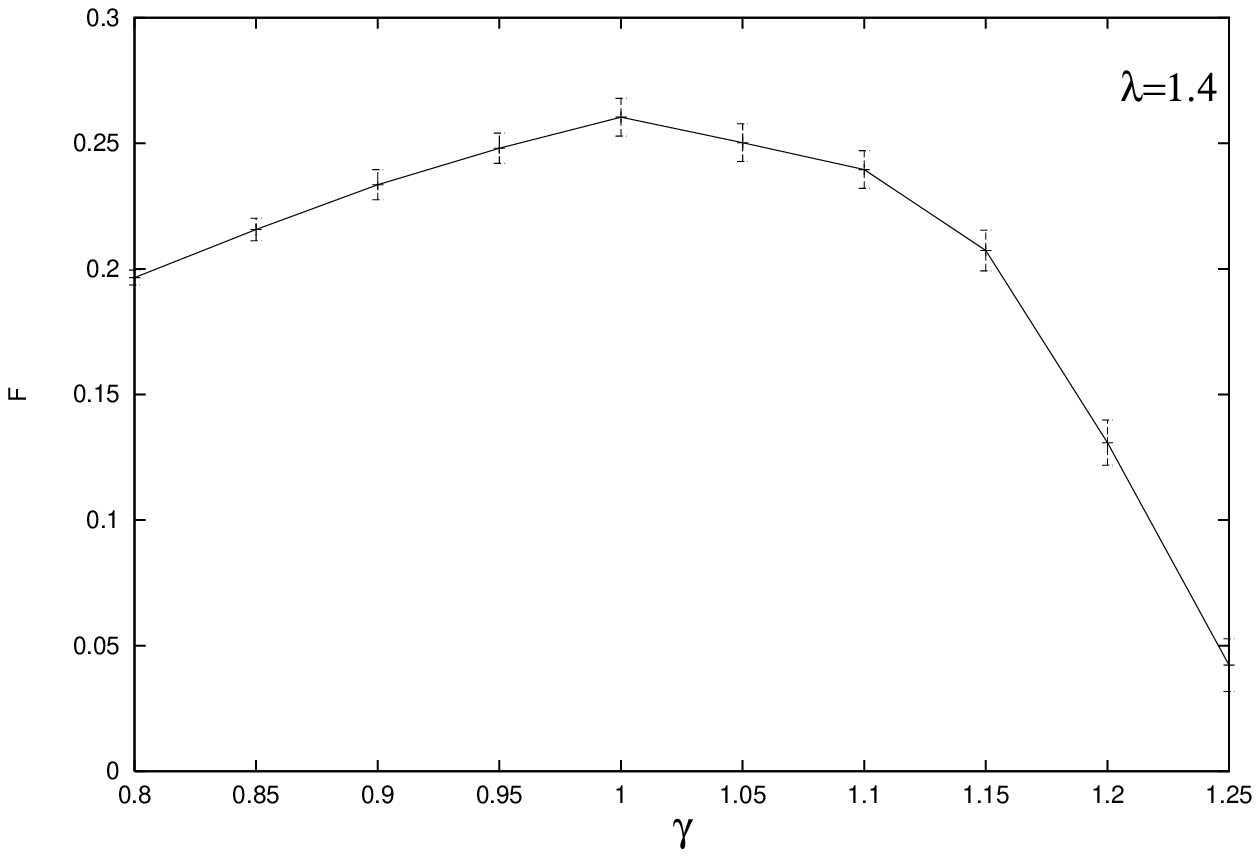}
  \end{tabular}
  \protect\caption{Plot of free energy $F$ vs. $\alpha$. The maximum
  value of $F$ is obtained for ($\alpha=1.0,\lambda=1.4$).}
  \label{MAXIMUM:FIG}
\end{center}
\end{figure}
The numerical simulations have clearly pointed out that the stationary
point of the functional is really a maximum confirming the
reasonment. Let us first present what happens for the coupling
constant $G=1$.
Setting $\lambda=1$ the free energy functional develops a maximum for
$\alpha=1.1$. The maximum value is obtained by varying also the
$\lambda$ parameter. Iterating this procedure in the interval $\lambda
\in\{0.8:1.3\}$ we find that as the value of $\lambda$ grows inside
the interval, the free energy maximum shifts from
$(\alpha=1.15,\lambda=1.0)$ to $(\alpha=1.0,\lambda=1.4)$ increasing
monotonically its value. 
Over the point $(\alpha=1.0,\lambda=1.4)$ 
we observe a decrease of functional. The range of the free energy
functional in this window of parameters belongs to the interval
$\{0.24:0.26\}$.\par
The value given in \formu{BETHE_ENERGY} is $F=1/3$ which is 
of the same magnitude of our simulation but significatively outside the
error bar. This difference from the analytic value must be addressed
to the fact that we are replacing a continuous model with a unitary step
discretization so that only wave functions which vary
on a length scale bigger than our discretization make sense. It is
therefore reasonable to introduce a modulation on the coupling between
replicas in \formu{ACCAVARn0} tuning the coupling constant $G$. 
The corresponding ground state energy is now $G^2/3$. As far as values
of $G<1$ are concerned, the binding between replicas is lowered, and
bound states with variations on a wider length scale are favored. We
can't consider anyway too small values of $G$ since in this case it's
impossible to build a bound state and the stationary wave function
turns out trivially to be constant. \par
\begin{figure}[!htb]
  \begin{center}
    \includegraphics[width=0.50\textwidth,height =0.25\textheight]{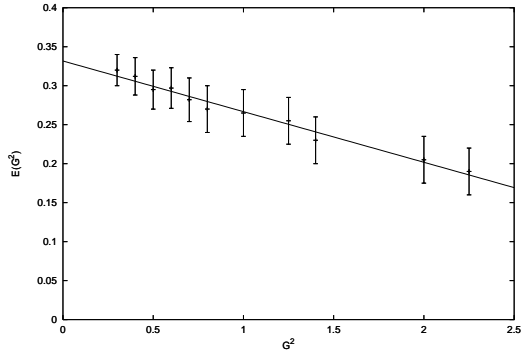}
    \protect\caption{Plot of $E(G)/G^2$ vs. the square of the coupling
    costant $G^2$. Here $N=64$ and $(\lambda=1.4,\alpha=1.0)$. From
    the linear fit extrapolate $\lim_{G^2\fd 0}E(G)/G^2=0.331\mp0.005$.}
    \label{EXTRAP:FIG}
  \end{center}
\end{figure}
Following the previous considerations for each value of the coupling
constant $G$ in the interval $\{0.5:1.2\}$ we have generated $10^5$
realizations of $\phi$ with $(\lambda=1.4,\alpha=1.0)$. As showed in
figure \formu{EXTRAP:FIG}
we were able extrapolate from the linear fit of $E(G)/G^2$ vs. $G^2$ the 
value of the abscissa $F=0.331\mp0.005$ which fits surprisingly well the
analytic value of $1/3$.\par
The variance around mean value overestimates the statistical error
related with our numerical estimates of free energy, since as we can
see from the normalized histogram of energy distribution in figure,
relevant power-law tails are present.
In order to give a reasonable estimate of  statistical errors, we
have generated for each experimental point $10^{6}$ realizations of
free energy, which we have randomly divided into $10$ subsets of $10^5$
elements. If - say $K$ - is the number of subsets, the statistical
estimate of variance made using the whole number of data we have, is
$K-1$ times bigger than the estimate given from averaging every
variance in the subsets built under the previous prescription.\par
It's worth noting that the variational parameter $\alpha$ is strictly
related with critical exponents of the model like for instance the
wandering exponent $\zeta$ defined by the relation
$\overline{\la|x|\ra}\sim t^\zeta$.
As pointed out in \cite{BO} $\phi(x)\sim x^2/t$, and since from
\formu{PHI_FLUCT} we know that $\phi(x)^2\sim x^\alpha$, is possible to
argue that $\zeta=2/(4-\alpha)$. In one dimension delta-correlated
potential we found the stationary point for $\alpha=1$, which means
$\zeta=2/3$, in complete agreement with Kardar's analytical solution
\cite{KARDAR_87}. 
\begin{figure}[!htb]
  \begin{center}
    \includegraphics[width=0.50\textwidth,height =0.25\textheight]{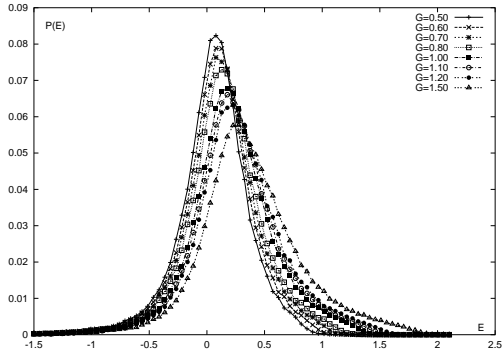}
    \protect\caption{Normalized histogram of the free energy frequency
    distribution for different $G$ obtained by $10^7$ iterations of
    \formu{MIO}. Here $N=64$, ($\alpha=1.0,\lambda=1.4$).}
    \label{HISTO:FIG}
  \end{center}
\end{figure}
\subsection{The power-law correlated potential}
The verification test implemented  for delta-correlated potential of
the variational principle is in good agreement with theoretical
predictions. Unfortunately no solution is available for more general
correlation of the random potential like power-law. It would be
interesting for instance to know how the wandering exponent depends on
power-law correlation's exponent. An approximate answer to this
question arise from the one loop renormalization group analysis
presented in \cite{MHKZ}, which lead to the following scenario 
(defining $\la
V(x)V(y)\ra\propto|x-y|^{-(2\rho-1)}\equiv|x-y|^{-\nu}$): 
\bi
\item[-]{ $\nu>1/2\;\;\;$ Renormalization group flux with one
stable point which scales to the delta-correlated potential behavior.
This regime is characterized by a fundamental length scale bigger then
the length scale over which the effects of the correlation of the
potential are sensible. Here $\zeta=2/3$ ($\alpha=1$) independently
from $\nu$. }
\item[-]{ $\nu<1/2\;\;\;$ In this regime correlations extends
on the same length scale of correlations and wandering exponent
depends on $\rho$. It is known sometimes as the Flory's regime and the
relation found is $\zeta(\nu)=3/(4+\nu)$ which following our notation 
also reads $\alpha(\nu)=(4-2\nu)/3$.}
\ei
\begin{figure}[!htb]
\begin{center}
  \begin{tabular}{rl}      
    \includegraphics[width=0.45\textwidth,height =0.25\textheight]{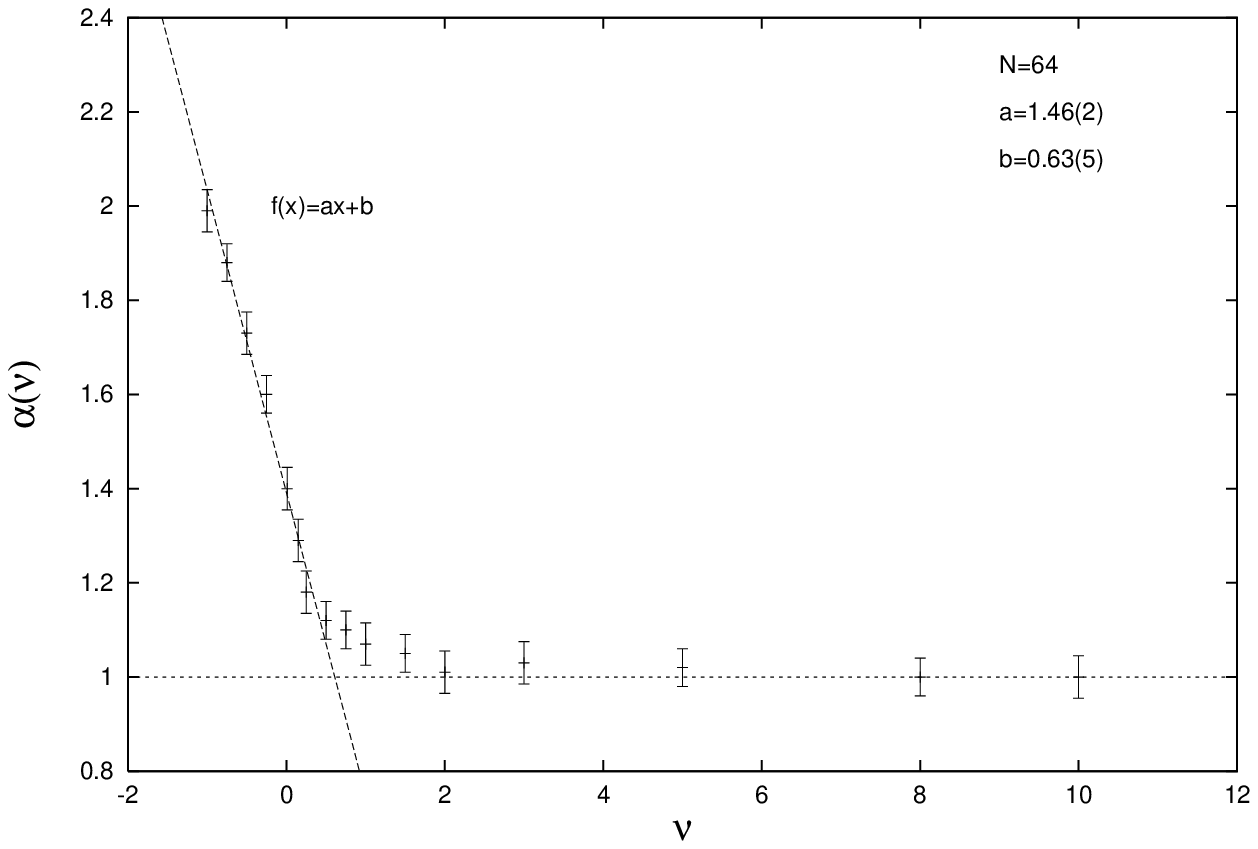}
   	&
    \includegraphics[width=0.45\textwidth,height =0.25\textheight]
	{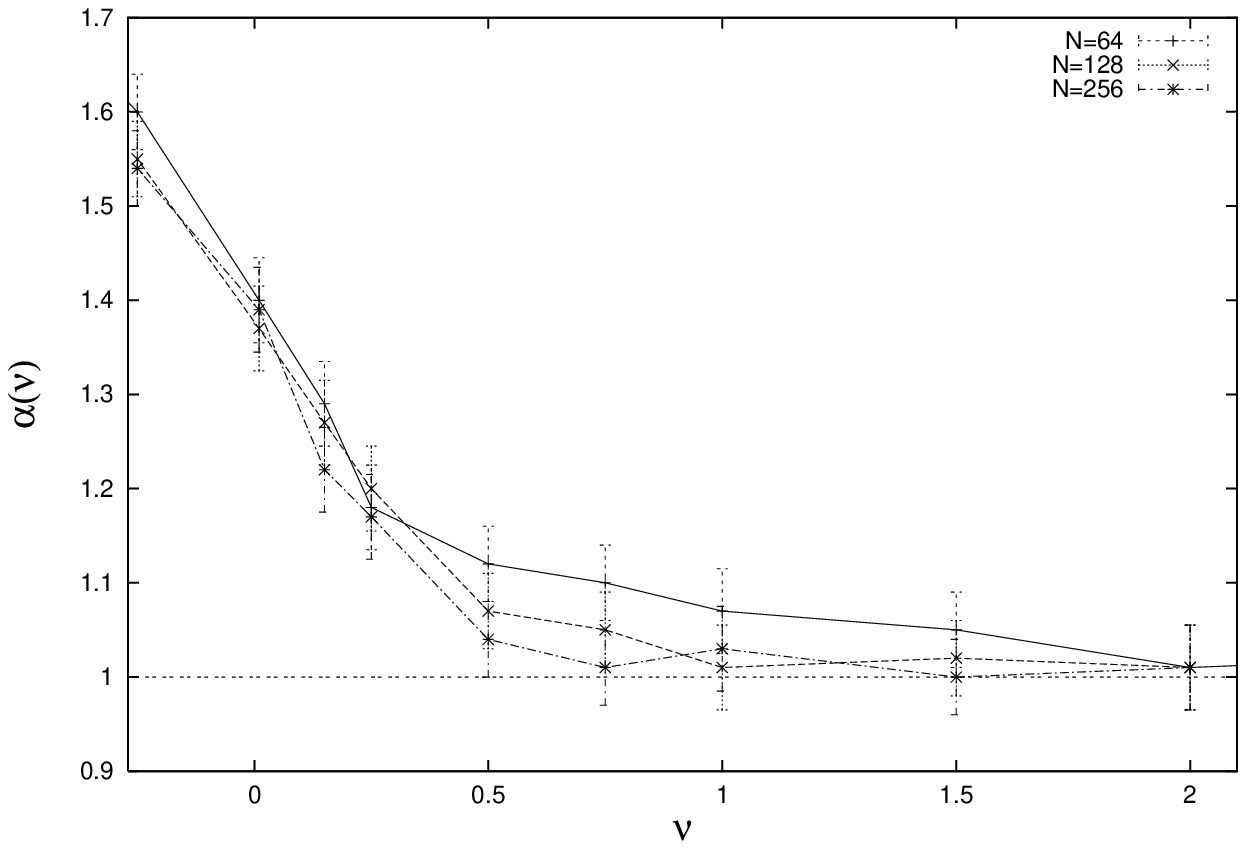}
  \end{tabular}
  \protect\caption{On the left is prented the behavior of parameter
	$\alpha$ which maximizes the functional vs. $\nu$ ($N=64$). 
	A transition to the Flory long-range scaling regime is clearly
	observable for $\nu>1/2$. On the left a zoom on the transition
	point is presented for $N=128,256,512$. We can observe a	
	more sharp transition from short to long range behavior as the
	volume increases.}
  \label{SCAL_NU:FIG}
\end{center}
\end{figure}
In order to verify the approximated predictions above, we have inserted
in the functional \formu{ACCAVARn0} a potentential of the following 
type:
\be             
V(x-y)=\frac{|x-y|^{-\nu}-1}{\nu}\;\;\;\;\;\;\;\nu=1-2\rho  
\label{VERY_NEW}
\ee
It has to be notice that functional structure of the correlations has
been chosen in order to obtain a logarithmic behavior in the limit
$\nu\fd 0$, so that formation of bound states is still allowed in
this limit. \par
The interval $\nu\in\{-2:10\}$ has been subjected to
investigation. Since negative values of the exponent $\nu$ obviously
correspond  to correlations which grow with the distance, the finite
size effects become more and more relevant. That's why was impossible
to analyze the behavior of the system for values of $\nu<-2$.\par
For each value of $\nu$ we have calculated the corresponding pair
$(\alpha,\lambda)$ which maximize \formu{VERY_NEW}, calculating free
energy and errors with the probabilistic prescription stated above.
The results are presented in figure (for details see caption). 
From figure \formu{SCAL_NU:FIG} 
we find a rather smooth behavior around the transition point
$\nu=1/2$. Iterating our analysis for polymers of different lengths
($N=128,256,512$) we can observe a more sharp transition since finite
size effects are less important. Figure \formu{SCAL_NU:FIG} shows that
for larger volume the different curves develop a real
discontinuity in the first derivative in $\nu=1/2$ in agreement with
the approximate results of \cite{MHKZ}.
\section{Conclusions}
The variational method developed seems to correctly describe
unidimensional polymers in random environment. For delta-correlated
potential, stationary point is effectively a maximum of our
replicated free energy functional in the $n\fd 0$ limit according with
replica theory. The value of the maximum free energy is in very good
agreement with Kardar's solution \cite{KARDAR_87}. \par
The power-law correlated potential has been subjected to
investigations and the results are in agreement with the perturbative
result in \cite{MHKZ}: for $\nu>1/2$ the wandering exponent equals the
delta-correlated one ($\zeta=2/3$), and for $\nu<1/2$ the standard Flory
regime is recovered ($\zeta(\nu)=3/(4+\nu)$).\par
A number of recently published papers pointed out that one
dimensional DPRM are characterized by replica symmetry breaking. Our
formulation can be in principle enriched introducing the replica
symmetry breaking as pointed out in
\cite{PARISI_TESI}. \par
Another development of the theory should be the analysis of
$d+1$-dimensional DPRM with $d>1$. Some preliminary tests performed in
$d=2$ pointed out that for the delta-correlated potential one 
must chose a probability distribution which is not a simple power-law in
$k$. Further inquiry are required maybe
in the direction of a more complex functional structure of equilibrium
probability distribution function. 
\\
\\
{\bf {Acknowledgments}}
I am deeply indebted with Professor Giorgio Parisi 
for his continuous support.

\end{document}